\documentclass[twocolumn,showpacs,preprintnumbers,amsmath,amssymb]{revtex4}

\usepackage{graphicx}
\usepackage{dcolumn}
\usepackage{bm}

\newcommand{\bq}{\begin{equation}}
\newcommand{\eq}{\end{equation}}
\newcommand{\bqa}{\begin{eqnarray}}
\newcommand{\eqa}{\end{eqnarray}}
\newcommand{\nn}{\nonumber \\}

\def\be     {\begin{equation}}
\def\ee     {\end{equation}}
\def\bea        {\begin{eqnarray}}
\def\eea        {\end{eqnarray}}
\def\bnn    {\begin{eqnarray*}}
\def\enn    {\end{eqnarray*}}

\begin{document}

\title{Violation of Wiedemann-Franz law
at the Kondo breakdown quantum critical point}
\author{K.-S. Kim$^{*}$ and C. P\'epin}
\affiliation{ Institut de Physique Th\'eorique, CEA, IPhT, CNRS,
URA 2306, F-91191 Gif-sur-Yvette, France} \affiliation{$^{*}$Asia
Pacific Center for Theoretical Physics, Hogil Kim Memorial
building 5th floor, POSTECH, Hyoja-dong, Namgu, Pohang 790-784,
Korea}
\date{\today}

\begin{abstract}
We study both the electrical and thermal transport near the
heavy-fermion quantum critical point (QCP), identified with the
breakdown of the Kondo effect as an orbital selective Mott
transition. We show that the contribution to the electrical
conductivity comes mainly from conduction electrons while the
thermal conductivity is given by both conduction electrons and
localized fermions (spinons), scattered with dynamical exponent $z
= 3$. This scattering mechanism gives rise to a quasi-linear
temperature dependence of the electrical and thermal resistivity.
The characteristic feature of the Kondo breakdown scenario turns
out to be emergence of additional entropy carriers, that is,
spinon excitations. As a result, we find that the Wiedemann-Franz
ratio should be larger than the standard value, a fact which
enables to differentiate  the Kondo breakdown scenario from the
Hertz-Moriya-Millis framework.
\end{abstract}


\maketitle

Existence of quasiparticles is the cornerstone in Landau's
Fermi-liquid theory \cite{LFL} for modern theory of metals. Since
they transport not only electric charge but also entropy, one sees
that the ratio ($L = \frac{\kappa}{T \sigma}$) between thermal
($\kappa$) and electrical ($\sigma$) conductivities is given by a
universal number $L_{0} = \frac{\pi^{2}}{3} \Bigl( \frac{k_{B}}{e}
\Bigr)^{2} = 2.45 \times 10 ^{-8} W \Omega K^{-2}$
\cite{Sommerfeld}, provided quasiparticles do not lose their
energy during collisions, and certainly satisfied at zero
temperature in the Landau Fermi liquid theory. Not only conventional metals
\cite{Ashcroft_Mermin} but also strongly correlated metals such as
heavy fermions \cite{Violation_WF_Optimal} turn out to follow the
Wiedemann-Franz (WF) law \cite{WF_Law}. In particular, even
quantum critical metals of $CeNi_{2}Ge_{2}$ \cite{CeNi2Ge2},
$CeRhIn_{5}$ \cite{CeRhIn5}, and $CeCoIn_{5}$ \cite{CeCoIn5} have
shown to satisfy the WF law at least in the zero temperature
limit, thus validating the quasiparticle picture, although their
resistivities deviate from the conventional $T^{2}$ behavior.

Several years ago, violation of the WF law was observed in the
optimally electron-doped cuprate $(Pr,Ce)_{2}CuO_{4}$
\cite{Violation_WF_Optimal} and hole-underdoped cuprate
$La_{2-x}Sr_{x}CuO_{4}$ \cite{Violation_WF_Under} while the WF law
turns out to hold in the overdoped cuprate
$Tl_{2}Ba_{2}CuO_{6+\delta}$ \cite{Violation_WF_Over}, suggesting
emergence of non-Fermi liquid physics as proximity of a Mott
insulator. Recently, anisotropic violation of the WF law has been
reported near the quantum critical point (QCP) of a typical
heavy-fermion compound $CeCoIn_{5}$, where only c-axis transport
violates the WF law while ab-plane transport follows it
\cite{HF_Violation_WF}. In this experiment the authors speculated
that the temperature quasi-linear electrical resistivity and
vanishing spectral weight may be one common feature for such
non-Fermi liquid physics. In this respect they naturally proposed
to observe violation of the WF law in $YbRh_{2}Si_{2}$ as another
typical heavy-fermion compound exhibiting similar non-Fermi liquid
physics with the c-axis measurement of $CeCoIn_{5}$, where both
ab- and c- axis transport show the temperature quasi-linear
electrical resistivity and vanishing spectral weight
\cite{YbRh2Si2}.

Physically, one can expect violation of the WF law as proximity of
Mott physics or superconductivity away from quantum criticality,
and as emergence of non-Fermi liquid physics near QCPs. In a Mott
insulator the presence of charge gap makes electrical conductivity
vanish, but gapless spin excitations can carry entropy, causing $L
> L_{0}$ while Cooper pairs transport electric currents without
entropy in the superconducting state, resulting in $L < L_{0}$. On
the other hand, entropy is enhanced near QCPs due to critical
fluctuations, and violation of the WF law is expected in
principle.


In this paper we examine thermal transport and violation of the WF
law based on the Kondo breakdown scenario
\cite{Pepin_KBQCP,Paul_KBQCP} as one possible heavy-fermion
quantum transition for $YbRh_{2}Si_{2}$. This scenario differs
from the standard model of quantum criticality in a metallic
system, referred as the Hertz-Moriya-Millis framework \cite{HMM},
in respect that in the former case the whole heavy Fermi surface
is destabilized at the QCP.

Several heavy-fermion compounds have been shown not to follow the
SDW theory \cite{YbRh2Si2,INS_Local_AF,dHvA,Hall}. Strong
divergence of the effective mass near the QCP \cite{dHvA} and the
presence of localized magnetic moments at the transition towards
magnetism \cite{INS_Local_AF} seem to support a more exotic
scenario. In addition, rather large entropy and small magnetic
moments in the antiferromagnetic phase may be associated with
antiferromagnetism out of a spin liquid Mott insulator
\cite{Senthil_Vojta_Sachdev}. Combined with the Fermi surface
reconstruction at the QCP \cite{dHvA,Hall}, this quantum
transition is assumed to show breakdown of the Kondo effect as an
orbital selective Mott transition
\cite{Senthil_Vojta_Sachdev,DMFT}, where only the f-electrons
experience the metal-insulator transition.

Our main result is that the WF law should be violated at the
Kondo breakdown QCP as proximity of spin liquid Mott physics, thus
$L > L_{0}$, resulting from the presence of additional entropy
carriers, here spinon excitations. This result may be opposed to the
preservation of the WF law in the SDW framework \cite{CeNi2Ge2}.

We start from the U(1) slave-boson representation of the Anderson
lattice model (ALM) in the large-$U$ limit \bqa && L_{ALM} =
\sum_{i} c_{i\sigma}^{\dagger}(\partial_{\tau} - \mu)c_{i\sigma} -
t \sum_{\langle ij \rangle} (c_{i\sigma}^{\dagger}c_{j\sigma} +
H.c.) \nn && + V \sum_{i} (b_{i}f_{i\sigma}^{\dagger}c_{i\sigma} +
H.c.) + \sum_{i}b_{i}^{\dagger} \partial_{\tau} b_{i} \nn && +
\sum_{i}f_{i\sigma}^{\dagger}(\partial_{\tau} +
\epsilon_{f})f_{i\sigma} + J \sum_{\langle ij \rangle} (
f_{i\sigma}^{\dagger}\chi_{ij}f_{j\sigma} + H.c.) \nn && + i
\sum_{i} \lambda_{i} (b_{i}^{\dagger}b_{i} + f_{i\sigma}^{\dagger}
f_{i\sigma} - 1) + NJ \sum_{\langle ij \rangle} |\chi_{ij}|^{2} .
\eqa Here, $c_{i\sigma}$ and $d_{i\sigma} =
b_{i}^{\dagger}f_{i\sigma}$ are conduction electron with a
chemical potential $\mu$ and localized electron with an energy
level $\epsilon_{f}$, where $b_{i}$ and $f_{i\sigma}$ are holon
and spinon, associated with hybridization and spin fluctuations,
respectively. The spin-exchange term for the localized orbital is
introduced for competition with the hybridization term, and
decomposed via exchange hopping processes of spinons, where
$\chi_{ij}$ is a hopping parameter for the decomposition.
$\lambda_{i}$ is a Lagrange multiplier field to impose the single
occupancy constraint $b_{i}^{\dagger}b_{i} + f_{i\sigma}^{\dagger}
f_{i\sigma} = N/2$, where $N$ is the number of fermion flavors
with $\sigma = 1, ..., N$.

Performing the saddle-point approximation of $b_{i} \rightarrow
b$, $\chi_{ij} \rightarrow \chi$, and $i\lambda_{i} \rightarrow
\lambda$, one finds an orbital selective Mott transition as Kondo
breakdown at $J \approx T_{K}$, where a spin-liquid Mott insulator
($\langle b_{i} \rangle = 0$) arises in $J > T_{K}$ while a
heavy-fermion Fermi liquid ($\langle b_{i} \rangle \not= 0$)
results in $T_{K} > J$
\cite{Senthil_Vojta_Sachdev,Paul_KBQCP,Pepin_KBQCP}. Here, $T_{K}
= D \exp\Bigl(\frac{\epsilon_{f}}{N \rho_{c}V^{2}}\Bigr)$ is the
single-ion Kondo temperature, where $\rho_{c} \approx (2D)^{-1}$
is the density of states for conduction electrons with the half
bandwidth $D$.

One can read the WF ratio in the mean-field approximation. In the
heavy-fermion phase it is given by $L = L_{0}$, representing a
Fermi liquid state of heavy quasiparticles with a large Fermi
surface. On the other hand, it becomes \bqa && L = L_{0} \Bigl(
\frac{t + J \chi}{t} \Bigr)^{2} \nonumber \eqa in the spin liquid
phase, where $t$  is the hopping of the conduction electrons and
$\chi$ is the spin liquid parameter. By contrast, in the  U(1)
slave-boson mean-field theory of the t-J Hamiltonian, \bqa &&
H_{tJ}^{MF} = \sum_{\langle ij \rangle} \Bigl\{NJ|\chi_{ij}|^{2} -
(t \delta + J \chi_{ij})f_{j\sigma}^{\dagger}f_{i\sigma} - H.c.
\Bigr\} , \nonumber \eqa where the holon field is replaced with
its mean-field value of $b_{i} = \sqrt{\delta}$ with hole
concentration $\delta$, one finds $L = L_{0} \Bigl( \frac{t \delta
+ J \chi}{t\delta} \Bigr)^{2}$ \cite{TJ_WF_Violation}, which
represents a strong violation of the WF law at the vicinity of the
insulating phase. This comparison tells us that the orbital
selective Mott transition in the ALM has milder violation of the
WF law than the single-band Mott transition although the
underdoped state of the t-J model may have similarity with the
fractionalized Fermi liquid \cite{Senthil_Vojta_Sachdev} of the
ALM.

Fluctuation-corrections are treated in the Eliashberg framework
\cite{Pepin_KBQCP}. The main physics is that the Kondo breakdown QCP
is multi-scale. The dynamics of the hybridization fluctuations is
described by $z = 3$ critical theory due to Landau damping of
electron-spinon polarization above an intrinsic energy scale
$E^{*}$, while by $z = 2$ dilute Bose gas model below $E^{*}$,
where $z$ is the dynamical exponent. The energy scale $E^{*}$
originates from the mismatch of the Fermi surfaces of the conduction
electrons and spinons, shown to vary from ${\cal O}(10^{0})$ $mK$
to ${\cal O}(10^{2})$ $mK$. Based on the $z = 3$ quantum
criticality, a recent study \cite{GR_KB} has fitted the divergent
Gr\"uneisen ratio with an anomalous exponent $0.7$.

Transport coefficients can be found from the following transport
equations \bqa && \vec{J}_{el}^{c,f,b} =
K_{0}^{c,f,b}(\alpha_{c,f,b}\vec{E} + \beta_{c,f,b} \vec{\epsilon}
- \vec{\nabla} \mu_{c,f,b}) + K_{1}^{c,f,b}
\Bigl(\frac{-\vec{\nabla}T}{T}\Bigr) , \nn && \vec{J}_{th}^{c,f,b}
= K_{1}^{c,f,b}(\alpha_{c,f,b}\vec{E} + \beta_{c,f,b}
\vec{\epsilon} - \vec{\nabla} \mu_{c,f,b}) + K_{2}^{c,f,b}
\Bigl(\frac{-\vec{\nabla}T}{T}\Bigr) . \nn \eqa
$\vec{J}_{el(th)}^{c,f,b}$ is an electric (thermal) current for
conduction electrons, spinons, and holons, respectively, and
$\vec{E}$, $\vec{\epsilon}$, $\mu_{c,f,b}$, and $T$ are an
external electric field, internal one, each chemical potential,
and temperature, respectively, where $\alpha_{c,f,b} = 1, 0, -1$
and $\beta_{c,f,b} = 0, 1, 1$. $K_{0}^{c,f,b}$, $K_{1}^{c,f,b}$,
and $K_{2}^{c,f,b}$ are associated with electrical conductivity,
thermoelectric conductivity, and thermal conductivity for each
excitation, respectively. Obtaining $\vec{\epsilon}$ from the
current constraint $\vec{J}_{el}^{f} + \vec{J}_{el}^{b} = 0$ with
$\mu_{c} = \mu_{f} - \mu_{b}$, and considering the open-circuit
boundary condition, we find physical response functions for
electrical conductivity $\sigma_{t}$, thermoelectric conductivity
$p_{t}$, and thermal conductivity $\kappa_{t}$, \bqa && \sigma_{t}
= \sigma_{c} + \frac{\sigma_{b} \sigma_{f} }{\sigma_{b} +
\sigma_{f} } , ~~~~~ p_{t} = p_{c} + \frac{ \sigma_{b} p_{f} -
\sigma_{f} p_{b} }{\sigma_{b} + \sigma_{f} } , \nn &&
\frac{\kappa_{t}}{T} = \frac{\kappa_{c}}{T} + \frac{\kappa_{f}}{T}
+ \frac{\kappa_{b}}{T} - \frac{(p_{b} + p_{f})^{2}}{\sigma_{b} +
\sigma_{f}} - \frac{p_{t}^{2}}{\sigma_{t}}  \eqa with
$\sigma_{c,f,b} \equiv K_{0}^{c,f,b}$, $p_{c,f,b} \equiv
K_{1}^{c,f,b}/T$, and $\kappa_{c,f,b} \equiv K_{2}^{c,f,b}/T$. One
can also derive this general expression from the path-integral
representation with the covariant derivative $\vec{D} =
\vec{\nabla} - i\vec{A}_{el} - i\vec{A}_{th}(i\partial_{\tau})$ in
the continuum approximation, where $\vec{A}_{el}$ and
$\vec{A}_{th}$ are external electromagnetic and thermal vector
potential fields, respectively. We note that this expression
reduces to that of the  t-J model, when contributions from  the conduction
electrons are neglected \cite{Ioffe_Larkin}.

It is straightforward to evaluate all current-current correlation
functions in the one loop approximation. We find \bqa &&
\sigma_{c}(T) = \mathcal{C}\rho_{c}v_{F}^{c2}\tau_{c,sc}^{b}(T) ,
\nn && \sigma_{f}(T) = \frac{\mathcal{C} \rho_{f}v_{F}^{f2}
}{[\tau_{f, sc}^{b}(T)]^{-1} + [\tau_{f, tr}^{a}(T)]^{-1}} , \nn
&& p_{c}(T) = \frac{\pi^{2}}{3} \frac{c_{F}}{\epsilon_{F}} T
\sigma_{c}(T)  , ~~~ p_{f}(T) = \frac{\pi^{2}}{3}
\frac{c_{F}}{\epsilon_{F}} T \sigma_{f}(T) , \nn &&
\frac{\kappa_{c}(T)}{T} = \frac{\pi^{2}}{3} \sigma_{c}(T) , ~~~
\frac{\kappa_{f}(T)}{T} = \frac{\pi^{2}}{3} \sigma_{f}(T)   \eqa
with $\mathcal{C} = \frac{N}{\pi} \int_{-\infty}^{\infty}{d y}
\frac{1}{(y^{2}+1)^{2}}$. In the electrical conductivity
$\rho_{c(f)}$ and $v_{F}^{c(f)}$ are density of states and Fermi
velocity for conduction electrons (spinons), respectively.
$\tau_{c(f),sc}^{b}(T) = [\Im \Sigma_{c(f)}(T)]^{-1}$ is the
scattering time due to $z = 3$ hybridization fluctuations, given
by \bqa && \Im \Sigma_{c(f)}(T > E^{*}) = \frac{ m_{b}V^{2}}{12\pi
v_{F}^{f(c)}} T \ln \Bigl( \frac{2T}{E^{*}} \Bigr) , \nn &&  \Im
\Sigma_{c(f)}(T < E^{*}) = \frac{ m_{b}V^{2}}{12\pi v_{F}^{f(c)}}
\frac{T^{2}}{E^{*}} \ln 2 , \nonumber \eqa where $m_{b} = (2N
V^{2} \rho_{c})^{-1}$ is the band mass for holons. Note that
hybridization fluctuations are gapped at $T < E^{*}$, resulting in
the Fermi liquid like correction. $\tau_{f, tr}^{a}(T) = \Bigl\{
\Bigl(\frac{k_{F}^{f}}{16\pi N}\Bigr) \gamma_{a}^{\frac{2}{3}}
T^{\frac{5}{3}} \Bigr\}^{-1}$ is  the transport time associated with $z
= 3$ gauge fluctuations, where $\gamma_{a} \approx \pi/v_{F}^{f}$
is the Landau damping coefficient for gauge fluctuations and
$k_{F}^{f}$ is the Fermi momentum of spinons. In the
thermoelectric coefficient $\epsilon_{F}$ is the Fermi energy for
conduction electrons, and $c_{F}$ is a geometrical factor, here
$c_{F} = 3/2$ for the spherical Fermi surface \cite{Mahan_Book}.

Several remarks are in order for each transport coefficient. An
important point is that the vertex corrections for scattering with
hybridization fluctuations can be neglected, a unique feature of
the two band model, resulting from heavy mass of spinons
\cite{Pepin_KBQCP,Paul_KBQCP}. This allows us to replace the
transport time with the scattering time for such a process. On the
other hand, vertex corrections for scattering with gauge
fluctuations turn out to be crucial, where infrared divergence of
the self-energy correction at finite temperatures is cancelled via
the vertex correction, giving rise to gauge-invariant \cite{Nambu}
finite physical conductivity \cite{YBKim}. As a result, the gauge
non-invariant divergent spinon self-energy $\Im \Sigma_{f}^{a}(T)$
in $\Im \Sigma_{f}^{b}(T) + \Im \Sigma_{f}^{a}(T)$ of the
conductivity expression is replaced with the gauge invariant
finite transport time $[\tau_{f, tr}^{a}(T)]^{-1}$. Both
irrelevance (hybridization fluctuations) and relevance (gauge
fluctuations) of vertex corrections can be also checked in the
quantum Boltzman equation study.

Both the thermoelectric and thermal conductivities are nothing but
the Fermi liquid expressions, where each fermion sector satisfies
the WF law. Although inelastic scattering with both hybridization
and gauge fluctuations may modify the Fermi liquid expressions
beyond the one-loop approximation, the WF law for each fermion
sector will be preserved at least in the zero temperature limit,
where such inelastic scattering processes are suppressed. One may
regard the WF law for each fermion sector as the most important
assumption in this paper.

Transport coefficients for holon excitations turn out to be much
smaller than fermion contributions, that is, $\sigma_{c}(T) \geq
\sigma_{f}(T) \gg \sigma_{b}(T)$, $p_{c}(T) \geq p_{f}(T) \gg
p_{b}(T)$, and $\kappa_{c}(T) \geq \kappa_{f}(T) \gg
\kappa_{b}(T)$ as clearly shown in Fig. 1, thus irrelevant.
Physically the dominance of fermion contributions can be understood
from an argument of density of states. Since there are many
states at the Fermi surface in the vacuum state, their
conductivities diverge in the clean limit as the temperature goes down
to zero. On the other hand, there are no bosons at zero
temperature, thus their conductivity vanishes when $T \rightarrow
0$.

\begin{figure}[t]
\vspace{3cm} \includegraphics{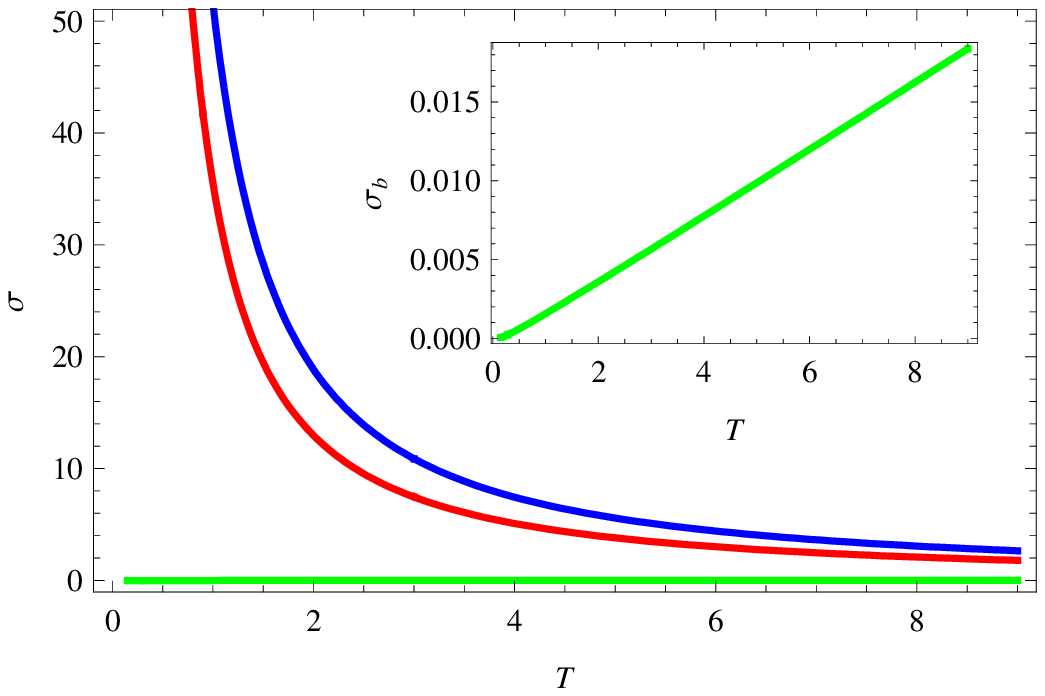}
\includegraphics{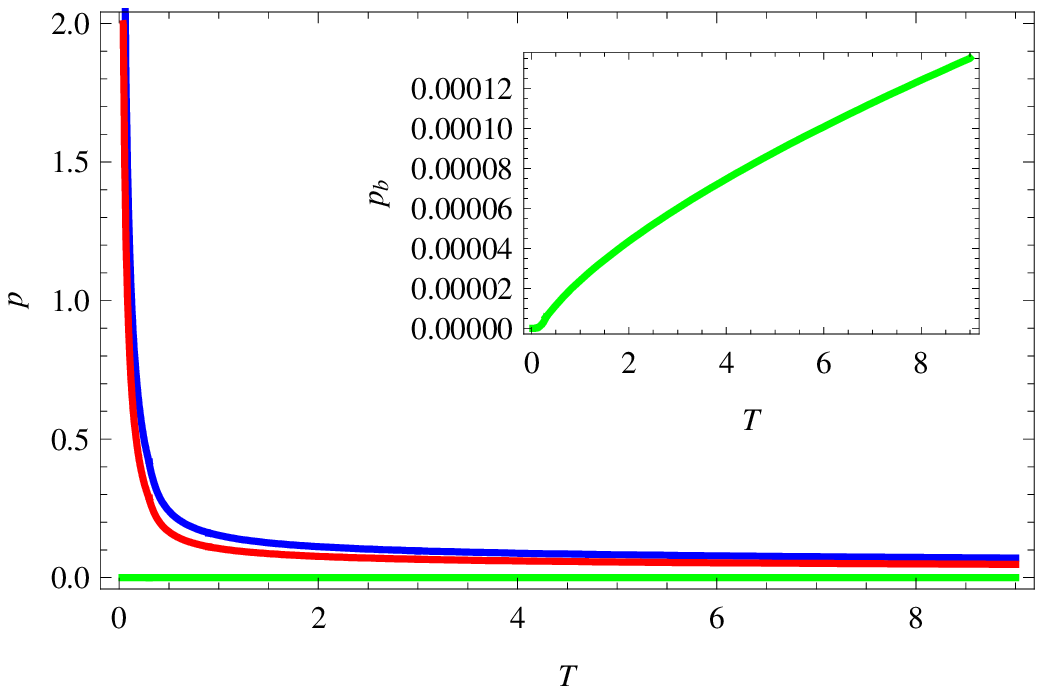} \vspace*{0cm} \vspace{-0.3cm}
\hspace{-1cm}
\hspace{3.7cm}
\caption{ (Color online) Left: Electrical conductivity from
conduction electrons (blue), spinons (red), and holons (green).
Left-inset: Electrical conductivity from holons much smaller than
contributions from fermions. Right: Thermoelectric conductivity
from conduction electrons (blue), spinons (red), and holons
(green). Right-inset: Thermoelectric conductivity from holons much
smaller than contributions from fermions.} \label{fig1}
\end{figure}

Inserting all contributions into Eq. (3), we find the physical
transport coefficients near the Kondo breakdown QCP.
Interestingly, the dominance of fermion contributions allows us to
simplify the total transport coefficients as \bqa && \sigma_{t}(T)
\approx \sigma_{c}(T) =
\mathcal{C}\rho_{c}v_{F}^{c2}\tau_{sc}^{c}(T) , \nn && p_{t}(T)
\approx p_{c}(T) = \frac{\pi^{2}}{3} \frac{c_{F}}{\epsilon_{F}} T
\sigma_{c}(T)  , \nn && \frac{\kappa_{t}(T)}{T} \approx
\frac{\kappa_{c}(T)}{T} + \frac{\kappa_{f}(T)}{T} =
\frac{\pi^{2}}{3} \Bigl( \sigma_{c}(T) + \sigma_{f}(T) \Bigr) .
\eqa Actually, we have checked that each approximate formula
matches with each total expression. The main point is that spinons
participate in carrying entropy, enhancing the thermal
conductivity, while both electric and thermoelectric
conductivities result from conduction electrons dominantly.

Fig. 2 shows the quasi-linear behavior in temperature for both
electrical and thermal resistivities above $E^{*}$, resulting from
scattering with $z = 3$ hybridization fluctuations dominantly,
because of $[\tau_{f, sc}^{b}(T)]^{-1} \gg [\tau_{f,
tr}^{a}(T)]^{-1}$ in the spinon conductivity, explicitly checked
from numerical analysis and temperature dependence, thus
$\sigma_{f}(T) \approx \mathcal{C} \rho_{f}v_{F}^{f2}\tau_{f,
sc}^{b}(T)$. Here, we have used parameters of Ref. \cite{GR_KB},
shown to be successful for fitting of Gr\"uneisen ratio.

\begin{figure}[h]
\vspace{3 cm} \includegraphics{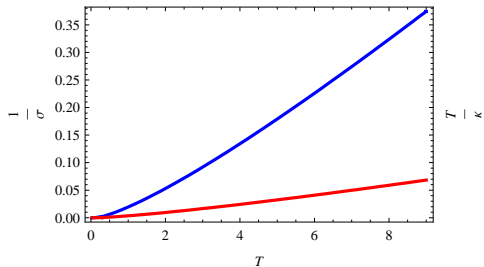} \caption{ (Color online)
Quasi-linear electrical (blue) and thermal (red) resistivity above
$E^{*}$, where thermal resistivity is smaller than electrical
resistivity owing to the contribution from spinon excitations.}
\label{fig2}
\end{figure}



The WF ratio is given by \bqa && L(T) \equiv
\frac{\kappa_{t}(T)}{T \sigma_{t}(T)} \approx \frac{\kappa_{c}(T)
+ \kappa_{f}(T)}{T \sigma_{c}(T)} \approx L_{0} \Bigl( 1 +
\frac{\rho_{f}v_{F}^{f}}{\rho_{c}v_{F}^{c}} \Bigr) ~~~~~~~  \eqa
in the low temperature limit, where gauge-fluctuation corrections
are irrelevant compared with hybridization-fluctuation
corrections, thus ignored in the last expression.


The larger value of the WF ratio is the characteristic feature of
the Kondo breakdown scenario, resulting from additional entropy
carriers, here the spinon excitations. If we perform the transport
study based on the SDW theory in the same approximation as the
present framework, we will find $\sigma_{t}(T) \propto
\tau_{tr}(T)$, $p_{t}(T) = \frac{\pi^{2}}{3}
\frac{c_{F}}{\epsilon_{F}} T \sigma_{t}(T)$, and
$\frac{\kappa_{t}(T)}{T} = \frac{\pi^{2}}{3} \sigma_{t}(T)$, where
likewise contributions from critical boson excitations are assumed
to be irrelevant, and the scattering time is replaced with the
transport time. As a result, the WF law is expected to hold
although non-Fermi liquid physics governs the quantum critical
regime. Actually, this has been clearly demonstrated in the
self-consistent renormalization theory, well applicable to
$CeNi_{2}Ge_{2}$ \cite{CeNi2Ge2}. In this respect the violation of
the WF law discriminates the Kondo breakdown scenario from the SDW
framework.

In this study we have ignored thermal currents driven by gauge
fluctuations, known to be the phonon-drag effect in the
electron-phonon system \cite{Mahan_Book}. Recently, Nave and Lee
have considered such photon-drag effects in the spin liquid
context with $z = 3$ gauge fluctuations based on quantum Boltzmann
equations, and argued that drag effects are subdominant, compared
with fermion contributions \cite{Nave_Lee}. Resorting to their
conclusion, we argue that dominant thermal currents are driven by
fermion excitations, here both conduction electrons and spinons.

In conclusion, we found marginal Fermi liquid physics for both
electrical and thermal transport near the Kondo breakdown QCP due
to scattering with $z = 3$ hybridization fluctuations. Our main
discovery is that the Kondo breakdown QCP should violate the WF
law at least in the zero temperature limit due to proximity of
spin liquid Mott physics, i.e., existence of additional entropy
carriers, that is, spinons.

We thank C. Bena and I. Paul for very helpful discussions. This work is
supported by the French National Grant ANR36ECCEZZZ. K.-S. Kim is
also supported by the Korea Research Foundation Grant
(KRF-2007-357-C00021) funded by the Korean Government.

\end{document}